\documentclass[aps,twocolumn,groupedaddress, showpacs,prb]{revtex4}
\usepackage{graphicx}%

\begin{document}
% You should use BibTeX and apsrev.bst for references
\bibliographystyle{apsrev}

%\preprint{}

\title{Layer-multiplicity as a community order-parameter}
% Optional argument for running titles on pages
%\title[]{}


% other informatio
\author{P. Fraundorf}
\affiliation{Physics \& Astronomy/Center for NanoScience, U. Missouri-StL (63121)}
\affiliation{Physics, Washington University (63110), St. Louis, MO, USA}
\email[]{pfraundorf@umsl.edu}
%\thanks{}


\date{\today}

\begin{abstract}

A small number of (perhaps only 6) broken-symmetries, marked by the edges of a hierarchical series of physical {\em subsystem-types}, underlie the delicate correlation-based complexity of life on our planet's surface. Order-parameters associated with these broken symmetries might in the future help us broaden our definitions of community health. For instance we show that a model of metazoan attention-focus, on correlation-layers that look in/out from the 3 boundaries of skin, family \& culture, predicts that behaviorally-diverse communities require a characteristic task layer-multiplicity {\em per individual} of only about $4 \frac14$ of the six correlation layers that comprise that community. The model may facilitate explorations of task-layer diversity, go beyond GDP \& body count in quantifying the impact of policy-changes \& disasters, and help manage electronic idea-streams in ways that strengthen community networks. Empirical methods for acquiring task-layer multiplicity data are in their infancy, although experience-sampling via cell-phone button-clicks might be one place to start.

\end{abstract}
\pacs{05.70.Ce, 02.50.Wp, 75.10.Hk, 01.55.+b}
%\keywords{}
\maketitle

\tableofcontents
\section{Introduction}

From the perspective of subsystem B, one might describe complete ignorance of an evolving subsytem A as perfectly symmetric since it attributes to A no special locations, directions, or excitations. Interactions that correlate subsystem B with an evolving subsystem A might provide information to B about specific locations, directions, and excitations in subsystem A, thereby breaking that perfect symmetry.

Gibb's dimensionless thermodynamic-availability\cite{Gibbs1873}, in modern terms known as Kullback-Leibler divergence i.e. mutual-information with respect to an arbitrary prior, is a measure of the correlation-information between subsystems A and B. The 2nd Law requires that our correlations with a subsystem A from which we are isolated can only decrease over time\cite{Lloyd89b}, but even then the time evolution of A's thermodynamic availability can give rise to the emergence in A of new symmetry breaks, or even a hierarchy of such breaks.

On the molecular level\cite{Ziman1979}, for instance, the relatively-featureless isotropic-symmetry of liquid water may on cooling first be broken by local translational pair-correlations (resulting in spherical reciprocal-lattice shells) as the liquid turns to polycrystal ice, and eventually by global translational and rotational ordering (resulting in reciprocal-lattice spots) as the ice becomes a single crystal. Partly along the way to single-crystal form a quasicrystal phase might have rotational without translational ordering, while a random-layer lattice might have rotational and translational ordering in one ``layering" direction only. Thus even within a single layer of organization, broken symmetries (often associated with a spatial gradient and/or boundary) play a role in the local development of order.

Complex systems often boast a hierarchical set of broken symmetries with associated gradients and/or boundaries. For instance a temperature-gradient marks the ``level-1" symmetry break that defines the center of a collapsing star system, within which local gravitational wells and condensed-matter surfaces associated with orbiting bodies (including planets) define ``level-2" symmetry-breaks.

In these gradients of our own planet a small number of (plausibly only six) additional broken-symmetries\cite{Anderson72}, again marked by the edges of a hierarchical series of physical subsystem-types\cite{pf.roots}, underlie the delicate correlation-based complexity of that interface-phenomenon that we call life. In this paper we explore how, by considering more than one level at a time, order-parameters\cite{Sethna2006} associated with these broken symmetries (which like standing-biomass and body-count are already quite useful) might help us broaden our definitions of community health\cite{pf.simplex}.

\section{Correlation-models}

In order to consider correlations on more than one level in hierarchical complex systems, we begin with ways to quantify pair and higher-order components of total-correlation (the generalization of mutual-information to more than two subsystems as a special-case of always-positive KL-divergence\cite{Kullback51}) on a single level. 

\begin{figure}
\includegraphics[scale=.90]{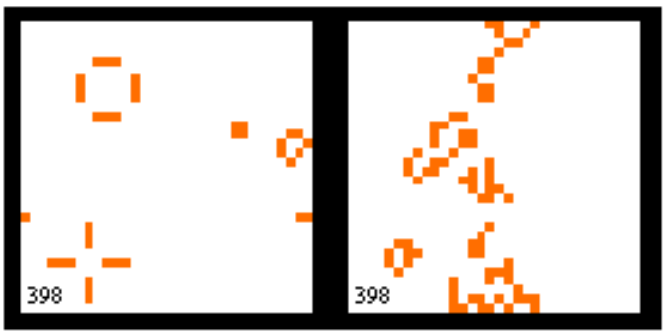}%
\caption{Generation 396 from a random start of Conway's life with 1 (left) and 0.99 (right) bits of state-information for each ``binary-state" neighbor, per generation.}
\label{Fig1}
\end{figure}

This may facilitate use of subsystem correlation-information as the natural thermodynamic limit on evolving complexity. As shown in Fig. \ref{Fig1}, for example, a limit on subsystem correlation-information in the John Conway's life adds realism to the process by preventing the steady-state endings to the evolution process.

Following Schneidman et al.\cite{Schneidman2003}, for example, we might say that the uncertainty associated with mth-order marginals for a system of L variables is something like:
\begin{equation}
S[\tilde{P}^{(m)}] =  \frac{(m-1)!(L-m)!}{(L-1)!} \sum_{i_{1}=1}^L \sum_{i_2>i_1}^L ... \sum_{i_m>i_{(m-1)}}^L S_{i_1 i_2 ... i_m}, 
\end{equation}
from which the "connected information" of order m for a system with L variables becomes, in terms of both our equations and their equation (6):
\begin{equation}
I_c^{(m)} = S[\tilde{P}^{(m-1)}] - S[\tilde{P}^{(m)}] \ge 0.
\end{equation}
Total correlation $I_c$ is simply the sum of these positive terms for $m$ running from 2 (pair correlations) up to L.

This unpacking of always-positive total-correlation measures into pair and post-pair components is of special interest to physicists because of the total-correlations connection, as a special case of KL-divergence, to applications for the second-law of thermodynamics. In fact the move to always-positive information-measures, like KL-divergence as the negative of Shannon-Jaynes entropy\cite{Gregory2005}, may signal a pedagogical move from entropy-1st thermodynamics to correlation-1st thermodynamics\cite{Fraundorf2011a} in the decades ahead.

\section{Multi-layer systems}

Observation of living systems on many levels, as well as of processes leading to planet formation and the biogeochemical cycles needed to support life, suggest that the establishment of subsystem-correlations may proceed inward and/or outward from a relatively-small number of very different emergent boundary types[6]. In the outward-looking case, development of subsystem-correlations often naturally starts with subsystem-subsystem i.e. pair interactions. Preliminary observations on single-level ordering in rather complex systems like neural nets\cite{Schneidman2006} also suggest that outward-focused ordering processes center primarily around subsystem pair correlations.

Boundary-emergence itself, and the post-pair or inward ordering processes subsequent thereto, are often treated as a separate subject. Traditional applications of long-standing interest here include e.g. studies of planetary accretion, homogeneous nucleation, and alliance-formation in context of the game-theory prisoner's dilemma\cite{Nowak2006}.

The presence of inward-looking subsystem-correlations is perhaps easiest to see once a higher-level of organization is in place. Thus for example mechanisms to preserve the integrity of a cell-membrane for controlled molecule-exchange with the outside world are likely in place partly to ensure microbe-viability, even though they manifest as multi-molecule correlations on the molecular level. Exploration of the precise mechanisms by which subsystem correlations at one level interact with subsystem correlations at the next level up is still in its early days, but of course is quite relevant e.g. to the topic of gene-pool/idea-pool co-evolution\cite{Richerson2004} so relevant in today's electronic-information age.

Thus hierarchical ordering in the layer just above a pair-correlated level (e.g. interacting organisms) may generally require a higher-level symmetry-break (e.g. recognition of differing organism groups), which in turn gives rise to processes that select for inward-looking (e.g. from the group boundary) post-pair correlations as well as outward-looking pair-correlations on the next level up (e.g. between groups).

Thus shared-electrons break the symmetry between in-molecule and extra-molecule interactions, bi-layer membranes allow symmetry between in-cell and out-cell chemistry to be broken, shared resources (like steady-state flows) may break the symmetry between in-tissue and external processes, metazoan skins allow symmetry between in-organism and out-organism processes to be broken, bias toward family breaks the symmetry between in-family and extra-familial processes, membership-rules break the symmetry between in-culture and multi-cultural processes, etc.

In this paper we focus on the perspective of (a) metazoan individuals as both audience and agent, instead of for instance on (b) the perspective of individual micro-organisms, or (c) the perspective of whole family gene-pools even though this is of much recent interest in biology. In that context, therefore, we center our attention on the last three symmetry-break levels (skin, family, culture) and the six subsystem-correlation layers associated therewith.

\section{A task layer-multiplicity simplex}

\begin{figure}
\includegraphics[scale=0.5]{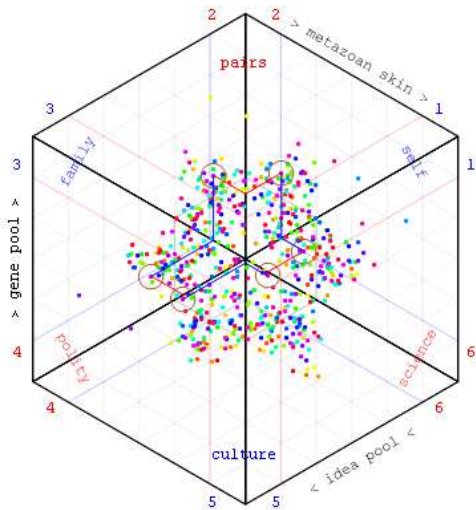}%
\caption{Six-projections of 100-member random simplex point-picked dot-cloud, with projections of one individual organism circled.}
\label{Fig2}
\end{figure}

Selection of order parameters for complex systems is sometimes more of an art than a science. Here as in the selection of order-parameters for simpler (albeit still-complex) thermodynamic systems[7], we seek a measure based on information available with minimal perturbation.

For inputs, we begin with six normalized positive numbers $f_i$ representing the fraction of an organism's effort allocated to each of the 6 subsystem correlation-layers i.e. which look in/out from skin, family and culture. For vizualization-purposes these six positive normalized $f_i$ values allow us to map the layer-focus of organisms to individual points within the {\em equilateral 5-simplex} between unit-vertices in 6-space (cf. Fig. \ref{Fig2}), just as ternary-diagrams map any three normalized positive-numbers onto an equilateral triangle or 2-simplex in 3-space. 

To inventory order we then define a single metazoan-individual's niche-network layer-multiplicity $m$ as the behavior-defined effective-number of correlation-buffering choices, expressed as an entropy-exponential in terms of that organism's set of 6 fractional-attention values $\{ f \}$:
\begin{equation}
1 \le \#_{\text{choices}} \equiv m[\{f\}] = \prod_{i=1}^{6}\left(\frac{1}{f_{i}}\right)^{f_{i}} = 2^{\#_{\text{bits}}} \le 6
\end{equation}
where $\Sigma_i f_i = 1$ i.e. sums to one over the level-index $i=1,6$. 

This multiplicity measure can also be expressed in terms of the number of bits of surprisal or state-uncertainty $S$ in bits about which correlation layer (e.g. self, friends, family, job, culture, profession) they are working on at any given time, i.e. $S = \ln_2 [m] = \Sigma_i f_i \ln_2 [1/f_i]$. However use of $\#_{\text{choices}}$ instead of $\#_{\text{bits}}$ probably makes more sense here since the numbers are so small.

Population-averages i.e. normalized-sums over all $N$ community members (say using index $j=1,N$) will be denoted with angle-brackets like $\langle \rangle$. Thus the {\bf population-average individual-multiplicity} is $\langle m \rangle = (1/N) \Sigma_j m_j$. The population-average value for attention-fraction $f_i$ is $\langle f_i \rangle = (1/N) \Sigma_j f_{ij}$ where $f_{ij}$ is the $j$th individual's layer $i$ attention-fraction. 

We'll use $\lbrace \langle f \rangle \rbrace$ to refer to the set of all 6 attention-fraction population-averages. This allows us to define a {\bf center-of-mass multiplicity} $M_{\text{cm}} = \Pi_i (1/{\langle f_i \rangle})^{\langle f_i \rangle}$, representing the spread in attention-focus for the community as a whole. 

We may also want to consider {\bf population average-surprisal} or entropy $\langle S \rangle = (1/N) \Sigma_j S_j$. This leads simply to the {\bf geometric-average individual-multiplicity}, defined as $M_{\text{geom}} = 2^{\langle S \rangle} = (\Pi_j m_j)^{1/N}$ for which it is easy to show that $M_{\text{geom}} \le M_{\text{cm}}$. Because of this organic relation to the center-of-mass value, we'll use $M_{\text{geom}}$ as our indicator of the spread in attention-focus for individual organisms with the community.

Finally, this inequality also lets us define organism and community {\bf specialization-indices}, whose logarithms are KL-divergences, which decrease in value toward 1 only as the spread of individual foci begins to match that of the community as a whole. For the community specialization index $R$, we use $1 \le R \equiv M_{\text{cm}}/M_{\text{geom}} \le M_{\text{cm}}$. For the $j^{\text{th}}$ individual organism the corresponding specialization index $r_j$ obeys $1 \le R \equiv m^*_j/m_j \le N$, where the individual center-of-mass multiplicity is defined as $m^*_j \equiv \Pi_i (1/\langle f_i \rangle)^{f_{ij}}$.

\section{Applications}

\begin{figure}
\includegraphics[scale=0.5]{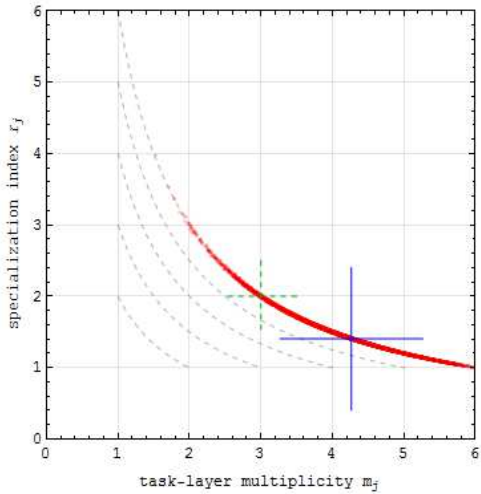}%
\caption{Task-layer specialization-index versus task-layer multiplicity for organisms in a 6-layer random simplex point-picked population of 10,000 individuals.}
\label{Fig3}
\end{figure}

We want to explore center-of-mass niche-network layer-multiplicity $M_{\text{cm}}$ as a measure of correlation-layer activity relevant to the survival of living systems, and the perhaps more subtle adaptive-value of task-layer diversity i.e. of a community with specialists and generalists of all sorts. These analyses treat all subsystem-correlation layers equally, in spite of a hierarchical structure which shows they are not. Let's begin therefore with their non-symmetric origins.

\subsection{evolving the hierarchy}

Imagine that $M_{\text{cm}}$ began increasing toward 2 when the metazoan skin of multi-celled organisms predicated the symmetry-break between self-focused behaviors (like hunger \& fear) and pair-focused behaviors (like aggression \& pair-bonding). When such social organisms began treating their young differently from the young of others, molecular code-pool boundaries facilitated the symmetry-break between family-focused behaviors (like bower-building \& child-rearing) and socially-focused behaviors (like status-pursuit \& community-service) letting $M_{\text{cm}}$ approach 4. $M_{\text{cm}}$ was allowed to approach 6 only after communicating organisms began recognizing distinctions between in-group and outsider patterns, allowing idea-pool symmetry-breaks to distinguish behaviors that are culturally-focused (like religion \& sports) and extra-cultural (like professional-development \& library-building). Astrophysical observations indicate that environments for such multi-layer correlation-structures are short-lived\cite{Ward03}, so quantitative models for $M_{\text{cm}}$'s increase \& decrease with time may be worthwhile.

\subsection{monitoring community health}

These models might provide integrative measures of social patterns already of interest, like division of responsibility between large and small gamete metazoans, and quantitative comparison of the extent and nature of community cultural-correlations from one species to another or from one time to another for a given species. If center-of-mass multiplicity correlates with other measures of health in human communities, it could be especially important for going beyond single-layer measures, like gross domestic product and body count, for taking quantitative account of family and culture when assessing the impact of policy changes and disasters on a given community.

\subsection{truth in advertising}

In addition to providing a window on the long-term activity-focus in a given community, dot clouds as in  Fig. \ref{Fig2} may help track the short-term effect of electronic media on the attention focus of a human community if real-time data is available e.g. via an electronic network. Related to this question is the use in electronic media of hooks to subconscious modules (like hunger, fear \& status-seeking) in the targeted population. If the effect of such hooks can be documented, the case for pointing out when they are being used might help temper their mis-use.

\subsection{task-layer diversification}

When task-diversity is maximized by random simplex point-picking, $M_{\text{cm}} \simeq 6$ but $M_{\text{geom}} \simeq 4.26$ i.e. everyone need not contribute on all layers. Is task layer diversity an adaptive feature of modern-day communities? 

The physiological division of labor between large and small gamete metazoans in reproductive roles shows that this may not always be the case. However communities with higher free-energy per capita and electronic information-flow seem to be moving away from cultural role-divisions. Fig. \ref{Fig3} illustrates by comparing $R$ and $M_{\text{geom}}$ of a 6-layer model with task-diversity maximized by random simplex point-picking (larger plus) with the same quantities for a ``yin-yang" community (smaller plus) in which half of the organisms each buffer subsystem correlations directed only inward, or only outward, from skin, family \& culture.

\section {The data challenge}

All of the applications above are predicated on a source of data about attention-focus in a given community. One may attempt to acquire data on some organism communities by direct observation. In human communities, self-reporting and communication-traffic analysis may also be useful particularly for data on short-term changes in attention-focus. A possible self-reporting strategy might for instance involve experience-sampling\cite{Hektner2007} by selecting a layer from 1 to 6 on your phone, when the occasional request comes in.

\section{Conclusion}

We describe in this paper a physical ``broken-symmetry'' approach toward community-structure inventories. It is integrative in that it is inspired by work on broken-symmetries in simpler physical systems, and in that its basics should apply to living systems on other levels of organization and in different astrophysical settings.

Note also that the discussion has not been about ``stand-alone states'' but about the relationship between subsystems, which subsystems of course include ourselves i.e. the knower. The promise of correlation-focused approaches to fundamental systems\cite{Markopoulou2000} may also, therefore, have practical benefits in the complex systems discussed here.

\begin{acknowledgments}
Thanks to Myron Tribus \& E. T. Jaynes for their work.
\end{acknowledgments}


%\begin{references}
\bibliography{ifzx2}
%\end{references}

\end{document}